\documentclass[
 reprint,onecolumn,notitlepage,
 amsmath,amssymb,
 aps,
12pt]{revtex4-2}

\usepackage{mathtools}

\usepackage{graphicx}
\usepackage{dcolumn}
\usepackage{bm}
\usepackage{xcolor}
\usepackage{hyperref}

\begin{document}


\title{Stability of cosmological singularity-free solutions in quadratic gravity } 

\author{Manuel Asorey\footnote{Electronic address: asorey@unizar.es}, Fernando Ezquerro\footnote{Electronic address: fezquerro@unizar.es} and Miguel Pardina\footnote{Electronic address: mpardina@unizar.es}}

\affiliation{%
Centro de Astropartículas y Física de Altas Energías (CAPA),
Departamento de Física Teórica. Universidad de Zaragoza,
E-50009 Zaragoza, Spain
}

\begin{abstract}
We introduce a large family of homogeneous and isotropic cosmological solutions in quadratic gravity which are singularity-free at early and late times. This kind of smooth solutions only emerges beyond the unstable de Sitter branch $3\alpha <  \beta$, $\alpha$ being the coupling of the $R^2$ term and $-\beta$ the coupling of the $R_{\mu\nu}^2$ term. We have analyzed the stability of these singularity-free solutions by computing the second-order variation of the action. The complete analysis shows that a Big Bang can emerge from a singularity-free regime when the parameters of the theory are slightly modified,  revealing the unstable nature of this type of solutions.
\end{abstract}

\keywords{quadratic gravity, cosmological solutions, dynamical stability}
\maketitle


\section{\label{sec:level1}Introduction}

It is well known that Einstein's General Relativity (GR) is a non-renormalizable theory from the viewpoint of Quantum Field Theory (QFT), since quantum corrections generate an infinite tower of extra counterterms that are not present in the original Einstein-Hilbert action and whose coefficients have to be fixed at every order of perturbation theory \cite{Goroff:1985th}. The addition to the Einstein-Hilbert action of quadratic terms in curvature tensors leads to a different gravitation theory, quadratic gravity, which is renormalizable \cite{Stelle:1976gc, Stelle:1977ry} and provides a ultraviolet (UV) completion of GR. Heuristically speaking, this important result follows from the fact that the curvature tensors are second order in derivatives and therefore the terms that are quadratic in the curvature are fourth order in derivatives, and the $1/k^4$ behavior of the corresponding propagator smooths out the ultraviolet behavior of the theory at high energies \cite{donoghue2021quadratic}. 

In spite of all these nice properties, quadratic gravity is still not free of pathologies. As pointed out by the pioneer work of Ostrogradsky theories with Higher Derivatives (HD) suffer from instabilities inducing a vacuum decay due the unbounded below character of corresponding Hamiltonian \cite{Woodard:2015zca}. 
This instability is commonly manifest at the quantum level as an extra degree of freedom, a ghost particle with negative kinetic term \cite{Asorey_1997, Sbis__2014}, which is typically associated in the literature with negative norm states and unitarity loss. There are recent and intense efforts to correctly understand this issue, but, although some progress has been made with different types of approach \cite{Asorey:2018wot, Kuntz:2019qcf, Donoghue_2021, PhysRevLett.128.041301, Mannheim:2021oat, deffayet2023global}, it remains a puzzling question. 

From a phenomenological viewpoint, quadratic gravity provided a pioneering model to describe the inflationary era of the primordial Universe \cite{Starobinsky:1980te}. This is a very conservative way to drive inflation, because there is no need of incorporating extra new degrees of freedom. After a Weyl transformation, the Starobinsky higher derivative term reveals by itself a scalar field minimally coupled to standard Einstein gravity with a potential that satisfies all inflationary requirements. It is remarkable that today this model is still one of the favorite candidates of the Planck collaboration \cite{Planck:2018jri}.

The aim of this paper is to analyze the cosmological implications of quadratic gravity, focusing on the fundamental differences with GR and proving that it is able of ameliorating the initial singularity behavior in some regions of the parameter space of the theory.

The structure of this paper is as follows: In section II we introduced the theoretical foundations of quadratic gravity.  The structure of the field equations restricted to a cosmological FLRW background metric is analyzed in Section III . In section IV we investigate families of different cosmological solutions for homegeneous and isotropic universes. Finally, the analysis of the stability of the singularity-free solutions of quadratic gravity is addressed in Section V. 

\section{Quadratic gravity action and field equations}

Quadratic gravity \cite{Schmidt:2006jt, Alvarez-Gaume:2015rwa, Salvio:2018crh} is the theory defined by adding all possible local quadratic curvature terms to the Einstein-Hilbert action, which can be expressed as
\begin{align}
    S &= \int d^4x \: \mathcal L = \int d^4x \sqrt{-g} \left[\gamma (R-2\Lambda) + \alpha R^2 -\beta R_{\mu\nu}R^{\mu\nu} \right] + S_m \; ,  \label{Sstelle}
\end{align}
where $\gamma=M_p^2/2$ is the usual coefficient of the Einstein-Hilbert action, 
$\Lambda$ is the cosmological constant, $S_m$ stands for the action corresponding to the matter content, $\alpha$ and $\beta$ are dimensionless coupling constants of the higher curvature terms. Under the general quadratic and local premise, we could also add to the action \eqref{Sstelle}, in principle, terms like  $R_{\mu\nu\alpha\beta}R^{\mu\nu\alpha\beta}$ or $\square R$. However, $\square R$ is a pure derivative  and in a four dimensional manifold $\mathcal M$, the integral of the Gauss-Bonnet term
\begin{align}
    G &= R^2 - 4 R_{\mu\nu}R^{\mu\nu} + R_{\mu\nu\alpha\beta}R^{\mu\nu\alpha\beta} \;  \label{GB}
\end{align}
is a topological invariant  $\chi(\mathcal M)=\frac{1}{32\pi^2}\int d^4x\sqrt{-g} 
\: G$ (Euler characteristic of the space-time manifold $\mathcal M$). This allows us to rewrite the Riemann-Riemann term $R_{\mu\nu\alpha\beta}R^{\mu\nu\alpha\beta}$ as a combination of the Ricci scalar squared $R^2$, the Ricci-Ricci term $R_{\mu\nu}R^{\mu\nu}$ and the Gauss-Bonnet invariant $G$, whose contribution is irrelevant if there is no topology change or boundary terms. On the other hand, the term $\sqrt{-g} \, \square R = \sqrt{-g}\, \nabla_\mu \nabla^\mu R = \partial_\mu\left( \sqrt{-g}\, \nabla^\mu R\right)$ is a total derivative and by Gauss's theorem can be written as an integral over the manifold's boundary,  thus we can ignore it under the same assumptions as before. Thus, adding $R_{\mu\nu\alpha\beta}R^{\mu\nu\alpha\beta}$ into the action \eqref{Sstelle} only produces a constant shift in the couplings and adding $\square R$ has no consequences for any observable which is insensitive to the overall topology of spacetime (such as the classical field equations or perturbative particle interactions \cite{Burgess_2004}). It is clear then that the theory described by the Stelle action \eqref{Sstelle} is the most general one for quadratic gravity.

  In the weak field limit, it can be shown \cite{Stelle:1977ry} that the resultant theory contains two extra gauge bosons in addition to the standard massless and spin 2 graviton from GR, associated with the $R$ term of the action. Indeed, the particle spectrum of the theory contains a spin 0 boson (the so-called scalaron) with mass $m_0=\sqrt{\frac{\gamma}{2(3\alpha-\beta)}}$, related to the $R^2$ term, and a spin 2 ghost with mass $m_2=\sqrt{\frac{\gamma}{\beta}}$, related to the $R_{\mu\nu}R^{\mu\nu}$ term. These massive modes are associated in the Newtonian limit with Yukawa-like potential behaviors, of the form $\frac{1}{r}e^{-m_0r}$ and $\frac{1}{r}e^{- m_2r}$, which smooth out the Newtonian singularity at the origin of a point-like source. In this limit, the linearized energy of the standard graviton and the massive scalaron is positive, while that of the massive spin 2 ghost is negative, as corresponds to its ghost nature. In the quantum regime, it is possible to obtain positive energy eigenstates for the ghost particle at the cost of dealing with negative norm states, violating in principle the unitarity of the theory.

The field equations derived from the action \eqref{Sstelle} are
\begin{align}
\gamma \left(G_{\mu\nu}+ \Lambda g_{\mu\nu}\right) + \Phi_{\mu\nu}  = \dfrac{1}{2} T_{\mu\nu} \; , \label{EcsCampoReducido}
\end{align}
where $T_{\mu\nu}$ is the energy-momentum tensor, $G_{\mu\nu}$ is the Einstein tensor and $\Phi_{\mu\nu}$ is the tensor associated to higher curvature terms (it can also be thought as an effective energy-momentum tensor quantifying the deviation from GR due to higher derivatives), defined by
\begin{align}
 T_{\mu\nu} &= - \dfrac{2}{\sqrt{-g}}\dfrac{\delta S_m}{\delta g^{\mu\nu}} \; , \label{Tmunu} \\
 G_{\mu\nu} &= R_{\mu\nu}- \dfrac{1}{2}g_{\mu\nu}R \; , \\
 \Phi_{\mu\nu} &= \alpha \left( 2RR_{\mu\nu} - \dfrac{1}{2}g_{\mu\nu}R^2 
- 
2\nabla_\mu  \nabla_\nu R  
+ 
2g_{\mu\nu} \square R \right)   \nonumber \\ 
& \hspace{0.4cm} - \beta\left(
- \dfrac{1}{2}g_{\mu\nu}R_{\alpha\beta}R^{\alpha\beta} - \nabla_\nu \nabla_\mu R
+
\square R_{\mu\nu}
+ 
\dfrac{1}{2} g_{\mu\nu} \square R
-2 R_{\alpha\mu\nu\beta}R^{\alpha\beta}
\right) \; .
\end{align}

Of course, when $\alpha=\beta=0$ then $\Phi_{\mu\nu}=0$, but the converse need not to be true. It is obvious that, in vacuum ($T_{\mu\nu}=0$) and ignoring cosmological constant ($\Lambda=0$), every solution to Einstein's equations $G_{\mu\nu}=0$ is also a solution of \eqref{EcsCampoReducido}. Furthermore, keeping the cosmological constant in vacuum shows that every maximally symmetric space in four dimensions is a solution of \eqref{EcsCampoReducido}. Indeed, in such a case
$T_{\mu\nu}=0$, $G_{\mu\nu}=-R_{\mu\nu}= -\Lambda g_{\mu\nu} $, $R=4\Lambda $ and it is straightforward to show that $\Phi_{\mu\nu}=0$.

\section{Field equations in a FLRW background metric}

Any solution of an homogeneous and isotropic  universe is of the Friedmann-Lema\^itre-Robertson-Walker (FLRW) metric form
\begin{equation}
    ds^2 = -b(t)^2dt^2 + a(t)^2(dx^2+dy^2+dz^2) \; , \label{FLRW}
\end{equation}
where usually the time component is reparametrized as $b(t)=1$ in GR. In that case, Einstein's equations for a perfect fluid provide the so-called two Friedmann equations, one associated to the time component and the other to the spatial components, which are linked by the local conservation law of matter $\nabla_\mu T^{\mu\nu}=0$ together with the state equation $p=\omega \rho$, relating the pressure $p$ with the energy density $\rho$ via the state parameter ($\omega=0$ for cold matter, $\omega=1/3$ for radiation and $\omega=-1$ for dark energy).

The situation in quadratic gravity is analogous. Of course, the problem can be solved by two equivalent but different formalisms: one proceeds by inserting  the metric \eqref{FLRW} and its associated curvature tensors into the general field equations \eqref{EcsCampoReducido} and the other is based on finding the stationary solutions of the effective action obtained by inserting the curvature tensors of the  FLRW metric \eqref{FLRW}  into the action \eqref{Sstelle} 
written in terms of the dynamic variables $a(t)$ and $b(t)$. Both procedures are completely equivalent, with the only difference that the metric is imposed before or after applying the variational principle to the action. The argument behind such equivalence is that there are only two physical degrees of freedom in massless gravity.

The curvature invariants of the quadratic action associated with the metric \eqref{FLRW} are given by 
\begin{align}
    R &= \dfrac{6}{b^2}(\Omega+\Theta) \; , \;\; R^2  = \dfrac{36}{b^4}\left(\Omega^2+\Theta^2+2\Omega\Theta \right) \; , \;\; R_{\mu\nu}R^{\mu\nu} = \dfrac{12}{b^4}\left(\Omega^2+\Theta^2+\Omega \Theta \right) \; , \nonumber \\
    &R_{\mu\nu\alpha\beta}R^{\mu\nu\alpha\beta} = \dfrac{12}{b^4}(\Omega^2+\Theta^2) \; , \;\; G = \dfrac{24}{b^4}\Omega\Theta \; , \; \; \text{where} \; \; \Omega=\frac{\ddot a}{a}-\frac{\dot a}{a}\frac{\dot b}{b} \; , \; \; \Theta = \frac{\dot a^2}{a^2} \; ,
    \label{curvatura}
\end{align}
where we denote $\dot x \equiv dx/dt$.
Thus the action, split into a gravitational $S_g$ part and a matter $S_m$ part is given by
\begin{align}
    S &= S_g+ S_m = \int d^3x \int dt \: \mathcal L(a,\dot a, \ddot a, b, \dot b) + S_m \; . \label{lag}
\end{align}
As it could be expected, the Lagrangian \eqref{lag} contains higher derivatives in the scale factor $a(t)$, but surprisingly only ordinary derivatives in the temporal component $b(t)$ are present. The corresponding Euler-Lagrange equations of the restricted action  are 
\begin{align}
   {\mathcal E}_b &= \dfrac{\partial \mathcal L}{\partial b} - \dfrac{d}{dt}\left(\dfrac{\partial \mathcal L}{\partial \dot b}\right) \; , \\ 
   {\mathcal E}_a &= \dfrac{\partial \mathcal L}{\partial a} - \dfrac{d}{dt}\left(\dfrac{\partial \mathcal L}{\partial \dot a}\right) + \dfrac{d^2}{dt^2}\left(\dfrac{\partial \mathcal L}{\partial \ddot a}\right)\; . \label{E-L a y b}
\end{align}
In vacuum, the equations of motion are ${\mathcal E}_b=0$ and ${\mathcal E}_a=0$. In the presence of a perfect matter fluid  the  energy-momentum tensor matter components are $T_{00}=b^2(t)\rho(t)$ and $T_{ij}=a^2(t)p(t)\delta_{ij}$.  Diffeomorphism invariance  implies  conservation laws of the Energy-Momentum tensor $\nabla^\mu T_{\mu0}=\dot\rho + 3\frac{\dot a}{a}(\rho+p)=0$ of the matter fluid 
\eqref{Tmunu}. The corresponding equations of motion are 
\begin{align}
   {\mathcal E}_b\left(a,\dot a, \ddot a, \dddot{a}, b, \dot b, \ddot b\right) &= \dfrac{\partial \mathcal L}{\partial b} - \dfrac{d}{dt}\left(\dfrac{\partial \mathcal L}{\partial \dot b}\right) = a^3\rho \; , \label{E-L b materia} \\ 
{\mathcal E}_a\left(a,\dot a, \ddot a, \dddot{a},\ddddot{a}, b, \dot b, \ddot b,\dddot{b}\right) &= \dfrac{\partial \mathcal L}{\partial a} - \dfrac{d}{dt}\left(\dfrac{\partial \mathcal L}{\partial \dot a}\right) + \dfrac{d^2}{dt^2}\left(\dfrac{\partial \mathcal L}{\partial \ddot a}\right) = -3a^2b p \; . \label{E-L a materia}
\end{align}
In view of the dependence of ${\mathcal E}_b$ and ${\mathcal E}_a$, it seems natural to derive ${\mathcal E}_b$ with respect to time to match the order of the system of differential equations in order to find a complete set of solutions of the Euler-Lagrange reduced equation. Now, the system composed by $\dot{{\mathcal E}}_b = 3a^2\dot a\rho + a^3\dot\rho $ and ${\mathcal E}_a=-3a^2bp$ has a  linear form
\begin{align}
\begin{rcases}
  A \ddddot a + B \dddot b &= P \\
  C \ddddot a + D \dddot b &= Q
\end{rcases} \; 
\begin{pmatrix}
A & B\\
C & D
\end{pmatrix} 
\begin{pmatrix}
\ddddot a \\
\dddot b
\end{pmatrix} = 
\begin{pmatrix}
P \\
Q
\end{pmatrix}
\end{align}
with the particularity that $\det M = AD - BC = 0$, indicating that the system is indeterminate and there is an infinity of solutions. This issue, which also occurs in GR and in any generally covariant theory, is resolved invoking the invariance under diffeomorphisms, which gives us the freedom to set a reparametrization of the temporal component, the simplest case being $b(t)=1$. The deep reason leans in the fact that temporal reparametrization is a subgroup of the group of diffeomorphisms that is a local gauge symmetry of generally covariant theories of gravitation and, therefore, all physical states and observables must be  invariant under diffeomorphisms. The gauge freedom allows us to set $b(t)=1$, making the system compatible with a unique solution. Under these conditions, the equations \eqref{E-L b materia} and \eqref{E-L a materia} can be explicitly written  as
\begin{align}
    a^3\rho &= \dfrac{12(3 \alpha -\beta )\dot{a}^2  \left(2 a \ddot a-3 \dot a^2\right)}{a}+6 a \left[\gamma  \dot a^2-2 (3 \alpha -\beta ) \left(\ddot a^2-2 \dddot a \dot a\right)\right]-2 \gamma  \Lambda 
   a^3 \; , \label{ec00} \\ 
   -3a^2 p &= 6\left[\gamma  \left(\dot a^2-\Lambda  a^2\right) + \dfrac{6 (3 \alpha -\beta ) \left(\dot{a}^4-4 a \dot{a}^2 \ddot{a}\right)}{a^2} + 2 a \left[2 (3 \alpha -\beta ) \ddddot a +\gamma  \ddot a\right] + \right. \nonumber  \\
   &\hspace{0.3cm}+ 
   2 (3 \alpha -\beta ) \left(3 \ddot a^2+4 \dddot a\dot a\right) \Biggr]   \; . \label{ecii}
\end{align}
 This quadratic gravity analogue of Friedmann equations of GR points out that both equations are not independent. They are related by matter equations and thus one of them is redundant. A straightforward computation shows that
\begin{align}
    \dot{\mathcal E}_b  = \dfrac{\dot a}{b}{\mathcal E}_a\; ,
\end{align}
which means that ${\mathcal E}_a$ is completely given in terms of ${\mathcal E}_b$. Since ${\mathcal E}_b$ has less higher derivative terms, it is the simplest of both equations and we choose it to solve the system of equations of motion.

Before solving equation \eqref{ec00}, it is worth to remark  that  in the special case $3\alpha=\beta$ between the higher derivative coupling constants  one recovers the ordinary Friedmann equations of GR from ${\mathcal E}_b$ and ${\mathcal E}_a$. The reason behind this relationship lies on the fact that the Ricci-Ricci term can be written as 
\begin{align}
    R_{\mu\nu}R^{\mu\nu} &= \dfrac{1}{3}R^2+\dfrac{1}{2}W^2 - \dfrac{1}{2}G \; , \label{relacioncurvaturas}
\end{align}
where $G$ is the Gauss-Bonnet invariant topological density \eqref{GB} and $W^2=W_{\mu\nu\alpha\beta}W^{\mu\nu\alpha\beta}$ is the curvature invariant associated with the Weyl tensor, $W_{\mu\nu\alpha\beta }$, which corresponds to the traceless components of the Riemann tensor. Moreover, for FLRW metrics \eqref{FLRW} the Weyl tensor vanishes identically \cite{Iihoshi_2007}, so $W^2=0$ and ignoring the topological term the classical action \eqref{Sstelle} reads
\begin{align}
    S &= \int d^4x \sqrt{-g} \left[\gamma (R-2\Lambda) + \left( 3\alpha-\beta\right)\dfrac{R^2}{3} \right] + S_m \; .  \label{SstelleModified}
\end{align}
It is clear now that the action \eqref{SstelleModified} is just the Einstein-Hilbert action when the higher derivative couplings satisfy $3\alpha=\beta$, and  in this case we recover the ordinary cosmological equations in this scenario. Furthemore, this procedure shows that quadratic gravity \eqref{Sstelle} restricted to FLRW metrics is completely equivalent to the $f(R)$ \cite{DeFelice:2010aj} Starobinsky-like model described by \eqref{SstelleModified}. 

If we write equation \eqref{ec00} in terms of the Hubble parameter $H=\dot{a}/a$ and in vacuum $\rho=0$ we find
\begin{align}
   & H^2 - \dfrac{\Lambda}{3}  + 2\dfrac{(3 \alpha -\beta)}{\gamma} \left( -\dot{H}^2 +6\dot H H^2 + 2 H\ddot H\right) =\frac{\rho}{6\gamma}
    = 0 \; , \label{Hec00} 
\end{align}
which manifests that de Sitter spacetime with $H=H_0=\sqrt{\Lambda/3}$ is a solution of the theory (as expected, because it is a maximally symmetric space). As was pointed out in \cite{TOPORENSKY_2007}, this de Sitter solution is stable if and only if $3\alpha>\beta$, i.e., when the coeficient of the $R^2$ term in \eqref{SstelleModified} is positive.

\section{FLRW cosmological solutions in quadratic gravity}

\begin{figure}[h]
	\begin{minipage}[c]{0.45\linewidth} 
		\scalebox{0.45}{\includegraphics{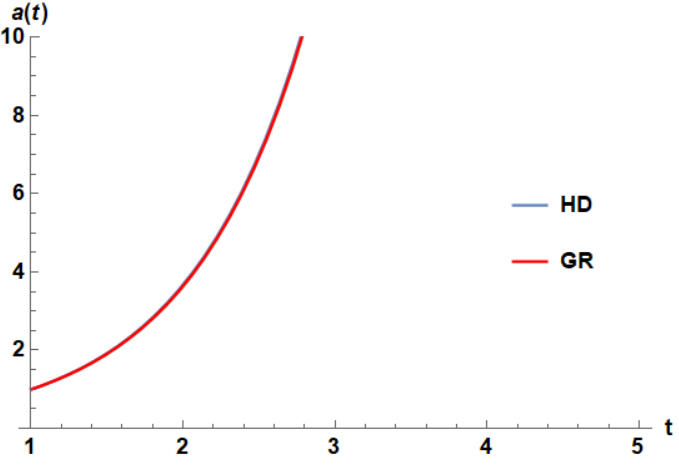}}
		\caption{Critical case, for parameters: $\Lambda=5$, $\alpha=1$, $\beta=2\text.9999$, $\gamma=20$, $a(1)=1$, $\dot a(1)=1$, $\ddot a(1)=0\text.2$.}\label{fig:critico}
	\end{minipage}
	\hspace{0.8cm}
	\begin{minipage}[c]{0.45\linewidth} 
		\scalebox{0.45}{\includegraphics{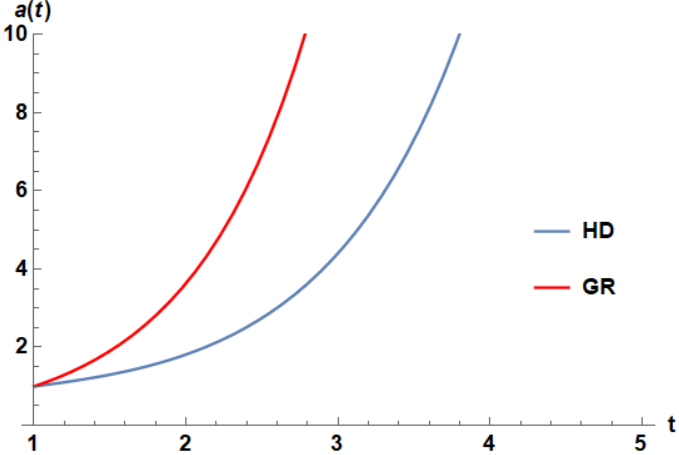}}
		\caption{Slower expansion, for parameters: $\Lambda=5$, $\alpha=5$, $\beta=1$, $\gamma=20$, $a(1)=1$, $\dot a(1)=0\text.5$, $\ddot a(1)=0\text.3$.}\label{fig:maslento}
	\end{minipage}
\end{figure}

\begin{figure}[h]
	\begin{minipage}[r]{0.45\linewidth}
		\scalebox{0.45}{\includegraphics{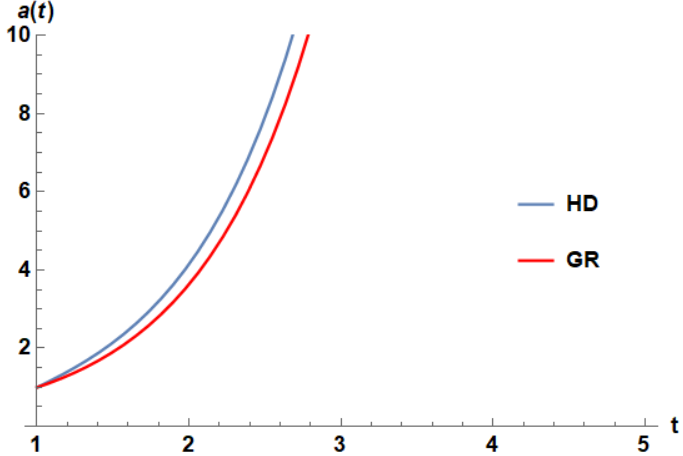}} have 
		\caption{Faster expansion, for parameters: $\Lambda=5$, $\alpha=5$, $\beta=1$, $\gamma=20$, $a(1)=1$, $\dot a(1)=1\text.9$, $\ddot a(1)=0\text.5$.}\label{fig:masrapido}
		\end{minipage}
		\hspace{0.8cm}
		\begin{minipage}[r]{0.45\linewidth}
		\scalebox{0.45}{\includegraphics{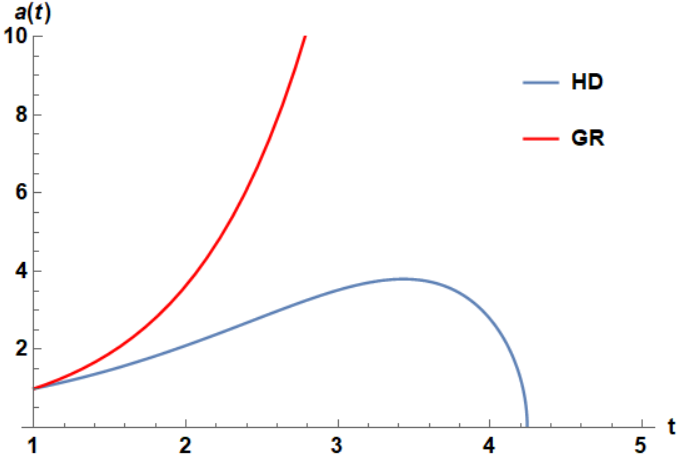}}
		\caption{Big Crunch, for parameters: $\Lambda=5$, $\alpha=5$, $\beta=30$, $\gamma=20$, $a(1)=1$, $\dot a(1)=0\text.8$, $\ddot a(1)=0\text.6$.}\label{fig:crunch}
		\end{minipage}
\end{figure}

If we solve equation \eqref{Hec00} in the stable regime $3\alpha>\beta$, we obtain a similar behavior to that of GR (which coincides with $3\alpha=\beta$, see \autoref{fig:critico}). This means that we have slower (\autoref{fig:maslento}) or faster (\autoref{fig:masrapido}) expansion rates for the Universe depending on the values of the parameters of the theory and the initial conditions, but all the solutions predict an expansion as the fate of the Universe. All diagrams show the cosmological  time evolution of scale factor, $a(t)$, for general relativity (red color, GR) and for quadratic gravity (blue color, HD).

Nevertheless, the unstable case given by $3\alpha<\beta$ is very interesting from a cosmological viewpoint. As can be seen in \autoref{fig:crunch} a new scenario emerges from this theory where the Universe eventually ends up collapsing ($a(t)\rightarrow 0$), regardless of the initial conditions. This behavior corresponds to a Big Crunch scenario. The result is interesting because it is a higher derivative solution that cannot occur in GR with such conditions, and it is essentially  a consequence of the relationship between the coupling constants of the higher curvatures. Furthermore, the unstable regime $3\alpha<\beta$ corresponds to imaginary values of the mass $m_0$ of the extra scalar particle. We remark  that in the presence of matter, none of the aforementioned behaviors changes substantially.

One of the major interests of theories beyond GR is to obtain a better understanding of the cosmological singularities that appears in Einstein's theory. In particular, it would be interesting to understand some features of the primordial initial singularity (Big Bang) from where the Universe emerges. Quadratic gravity has something to say about this scenario, and when the initial time behavior is analyzed we obtain a wide variety of solutions (see Figures 5-8)  .
\begin{figure}[h]
	\begin{minipage}[c]{0.45\linewidth} 
		\scalebox{0.45}{\includegraphics{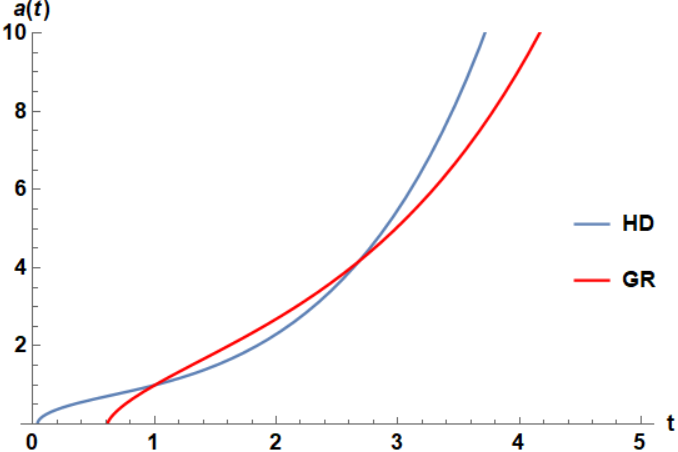}}
		\caption{Time evolution of the scale factor $a(t)$ in a stable matter  regime, parameters: $\omega=0$, $\rho_0=170$, $\Lambda=1$, $\alpha=4$, $\beta=3$, $\gamma=10$, $a(1)=1$, $\dot a(1)=0\text.8$, $\ddot a(1)=0\text.6$.}\label{fig:materia}
	\end{minipage}
	\hspace{0.8cm}
	\begin{minipage}[c]{0.45\linewidth} 
		\scalebox{0.45}{\includegraphics{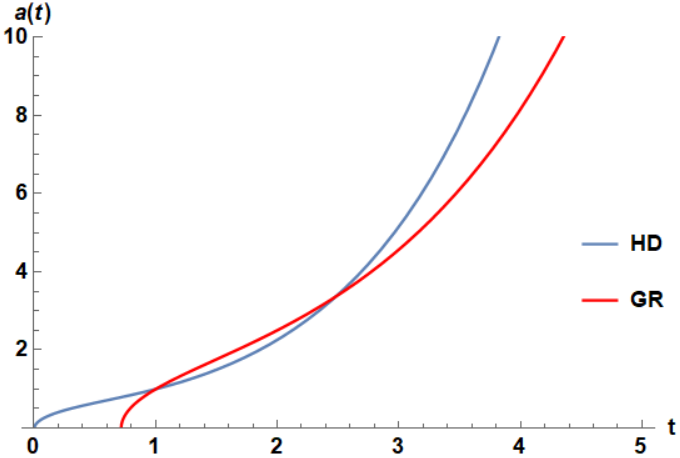}}
		\caption{Time evolution of the scale factor $a(t)$ in a stable radiation  regime, parameters: $\omega=1/3$, $\rho_0=180$, $\Lambda=1$, $\alpha=6$, $\beta=3$, $\gamma=10$, $a(1)=1$, $\dot a(1)=0\text.8$, $\ddot a(1)=0\text.6$.}\label{fig:rad1}
	\end{minipage}
\end{figure}
\begin{figure}[h]
	\begin{minipage}[r]{0.45\linewidth}
		\scalebox{0.45}{\includegraphics{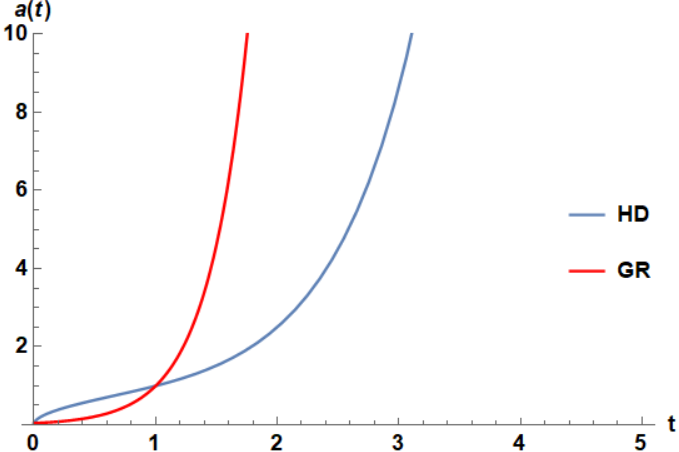}}
		\caption{Time evolution of the scale factor $a(t)$ in a stable dark energy regime, parameters: $\omega=-1$, $\rho_0=540$, $\Lambda=1$, $\alpha=8$, $\beta=4$, $\gamma=10$, $a(1)=1$, $\dot a(1)=0\text.8$, $\ddot a(1)=0\text.6$.}\label{fig:darkenergy}
		\end{minipage}
		\hspace{0.8cm}
		\begin{minipage}[r]{0.45\linewidth}
		\scalebox{0.45}{\includegraphics{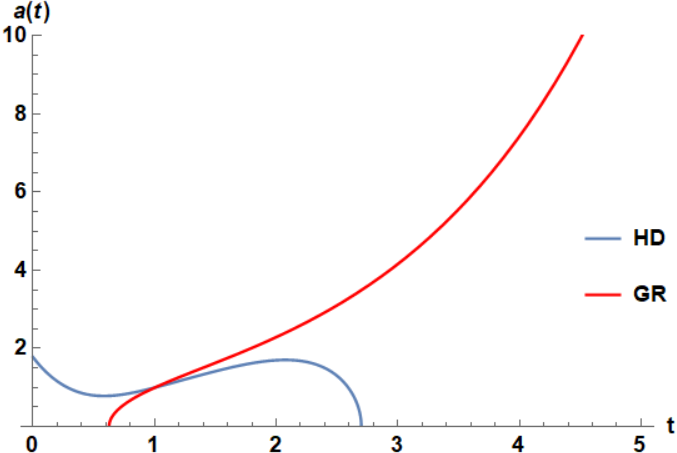}}
		\caption{Time evolution of the scale factor $a(t)$ in a  unstable radiation regime, parameters: $\omega=1/3$, $\rho_0=100$, $\Lambda=1$, $\alpha=5$, $\beta=16$, $\gamma=10$, $a(1)=1$, $\dot a(1)=0\text.8$, $\ddot a(1)=0\text.6$.}\label{fig:crunch2}
		\end{minipage}
\end{figure}

For instance, a scenario as the one displayed in \autoref{fig:materia} can appear in a Universe dominated by matter or the one displayed \autoref{fig:rad1} in a radiation dominated Universe or that of \autoref{fig:darkenergy} in an Universe dominated by dark energy. In all these cases of the stable regime $3\alpha>\beta$, there exists a Big Bang where the Universe is born, and the scale factor behaviors are similar to those of GR with cosmological constant. However, in \autoref{fig:crunch2}, which represents the unstable regime $3\alpha<\beta$, we obtain that the Universe can start from either a Big Bang or from infinite size to end collapsing into a Big Crunch regardless of the matter content considered. 

We remark that homogeneous, isotropic solutions of quadratic gravity  with initial singularity such as those shown in \autoref{fig:materia}, \autoref{fig:rad1} and \autoref{fig:darkenergy} (which also all behave as a radiation-dominated Friedmann Universe, $a(t)\sim t^{1/2}$ when $t\rightarrow 0$) are stable in presence of small scalar, vector and tensor heterogeneity due to the effect of the quadratic terms that drive the time evolution when approaching to the initial singularity \cite{Middleton:2008rh}. This is different from what happens in GR, where an instability occurs as $t\rightarrow 0$, and the same happens if general powers or functions of the Ricci-Ricci term are considered, indicating that the behavior of quadratic gravity is quite special. 

\begin{figure}[h]
	\centering	\scalebox{0.5}{\includegraphics{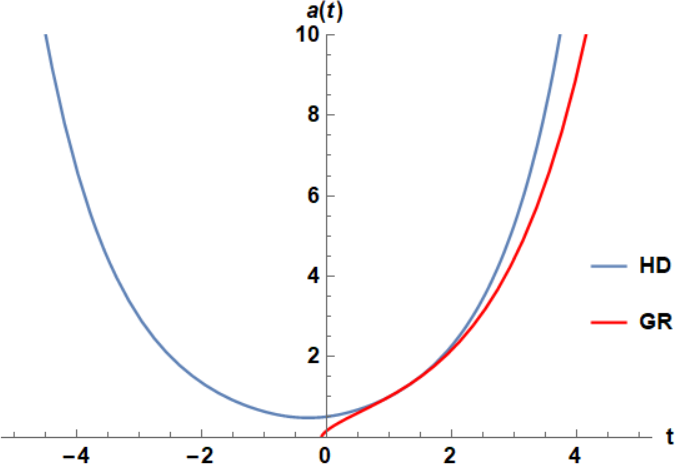}}
		\caption{Singularity-free solution of motion equations corresponding to the choice of parameters: $\omega=0$, $\rho_0=15$, $\Lambda=1\text.5$, $\alpha=3$, $\beta=16$, $\gamma=10$, $a(1)=1$, $\dot a(1)=0\text.8$, $\ddot a(1)=0\text.7$.}\label{fig:nosing}
\end{figure}

Furthermore, the cosmological landscape of scenarios in quadratic gravity is so rich that it is also possible to find solutions with an expansion at late times and without a  Big Bang singularity in early times, that is, solutions which are free of singularities. This behavior corresponds to a Universe that contracts to a finite minimum size and then expands again without limit. An example of this type of solution, which only occurs in the unstable branch $3\alpha<\beta$ in presence of matter (unlike what happened in late times, different behaviors do emerge when matter is incorporated in primordial times) is shown in \autoref{fig:nosing}. 

In Reference \cite{Kuntz:2019lzq} it was argued by analyzing Raychaudhuri's equation, that the existence of a singularity-free solution in quadratic gravity can be interpreted as a consequence of the ghost particle, whose classical repulsive energy could be able to prevent the formation of singularities in the spacetime. This is totally consistent with our analysis, since this kind of solutions appear when the $\beta$ coupling of the Ricci-Ricci term (responsible of the ghost particle) is larger than $3\alpha$.

Several studies have tried to promote this behavior to a periodic regime, joining contractions and expansions of the Universe continuously (the so-called big bounce or Ekpyrotic Universe \cite{Khoury:2001wf}). However, it is not clear how such a behavior can arise in this theory, since the semiclassical approximations used are no longer valid in the bounce \cite{Miranda_2022}.

\section{Stability of the nonsingular solutions}

It is also possible to answer the question of whether nonsingular solutions are stable under homogeneous perturbations are stable or not. The answer of the question requires the computation of  the second order variation of the action with respect to the dynamical fields $a(t)$, $\dot a(t)$ and $\ddot a(t)$, yielding to the general quadratic form 
\begin{align}
24\begin{pmatrix}
\frac{\gamma}{2}  \ddot a+\frac{(3 \alpha -\beta ) \dot a^4}{a^3}-\frac{\rho_0\omega (3\omega+1)}{8} a^{-3 \omega-2}- \frac{\gamma}{2}  \Lambda  a & \dot a \left(\frac{2(\beta -3 \alpha ) \dot a^2}{a^2}+\frac{\gamma}{2} \right) & (3 \alpha -\beta ) \ddot a+\frac{\gamma}{2}  a \\
\dot a \left(\frac{2 (\beta -3 \alpha ) \dot a^2}{a^2}+\frac{\gamma}{2} \right) & (6 \alpha -\beta ) \ddot a+\frac{6 (3 \alpha -\beta ) \dot a^2}{a}+\frac{\gamma}{2}  a &  (6 \alpha -\beta )\dot a \\
(3 \alpha -\beta ) \ddot a+\frac{\gamma}{2}a &  (6 \alpha -\beta ) \dot a &  (3 \alpha -\beta ) a \\
\end{pmatrix} \label{hessian}
\end{align}
where we suppressed the explicit time dependence for aesthetic reasons. The stability analysis now follows from the eigenvalues and eigenvectors of the spectrum of the quadratic form associated to the matrix (\ref{hessian}) in a background non-singular metric solution of equations of motion. 

\begin{figure}[h]
	\centering	\scalebox{0.5}{\includegraphics{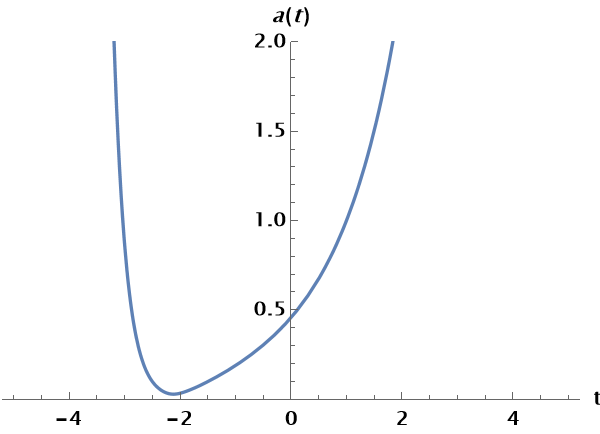}}
		\caption{Nonsingular solution for parameters: $\omega=0$, $\rho_0=1.66$, $\Lambda=0\text.5$, $\alpha=4$, $\beta=19$, $\gamma=10$, $a(1)=1$, $\dot a(1)=0\text.8$, $\ddot a(1)=0\text.7$.}\label{fig:nosingestable}
\end{figure}

It turns out that the behavior of the family of non-singular solutions is qualitatively the same, thus, it is sufficient to consider one particular case. We choose the solution shown in \autoref{fig:nosingestable}, whose parameters are close enough to the transition from a regular solution to a Big Bang solution, where the scale factor $a(t)$ crosses the horizontal axis and becomes zero.

Computing the second variation of the action \eqref{hessian} for such a solution shown above and diagonalizing the resulting matrix provides the following time dependent eigenvalues.
\begin{figure}[h]
	\centering	\scalebox{0.4}{\includegraphics{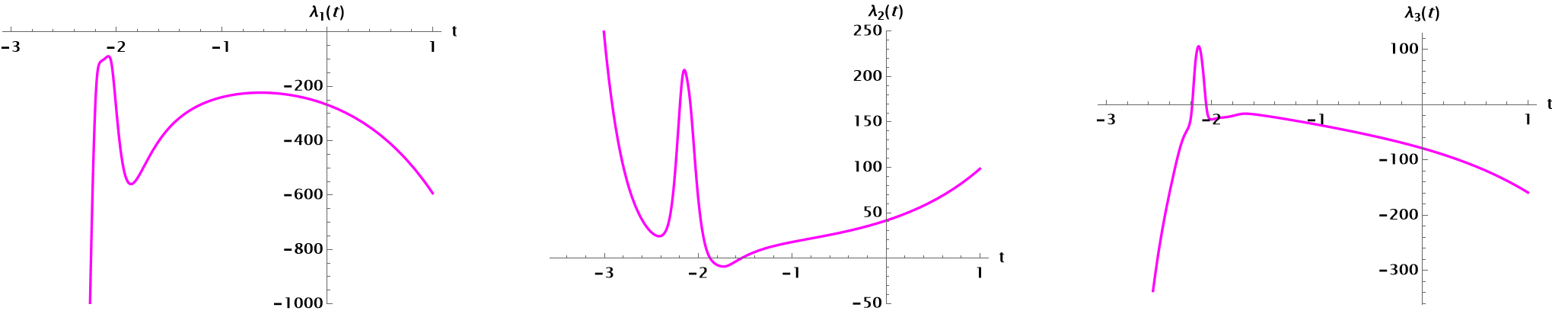}}
		\caption{Eigenvalues $\lambda_1(t), \lambda_2(t), \lambda_3(t)$ associated to the second order variation of the action for the solution shown in \autoref{fig:nosingestable}.}\label{fig:lambdas}
\end{figure}\\
As expected, the three eigenvalues show a quasi-pathological behavior at the same time value $t_c\approx 2\text. 12$, which is precisely the time where the scale factor (\autoref{fig:nosingestable}) presents a minimum near zero, $a(t_c)\approx 0\text. 028$. Indeed, the vertical asymptote of $\lambda_1(t)$ and the spiked shapes of $\lambda_2(t)$ and $\lambda_3(t)$ around $t=t_c$ are a consequence of the of the vanishing of the scale factor $a(t_c)$, \textit{i.e.}, to a singular Big Bang regime coming from a transition of a non-singular Universe. This analysis, together with the fact that the quadratic form is not positive defined as can be seen from the eigenvalues plot (\autoref{fig:lambdas}), suggests the existence of an  instability of the non-singular solutions considered. It is interesting to remark that the eigenvalue that drives the largest instability mode is $\lambda_1(t)$, thus,  the associated eigenvector $\bm v_1(t)=(v_{1x}(t),v_{1y}(t),v_{1z}(t))$ signals the direction of maximum instability of the system. The  components of this normalized eigenvector are shown in \autoref{fig:autovector1}.
\begin{figure}[h]
	\centering	\scalebox{0.4}{\includegraphics{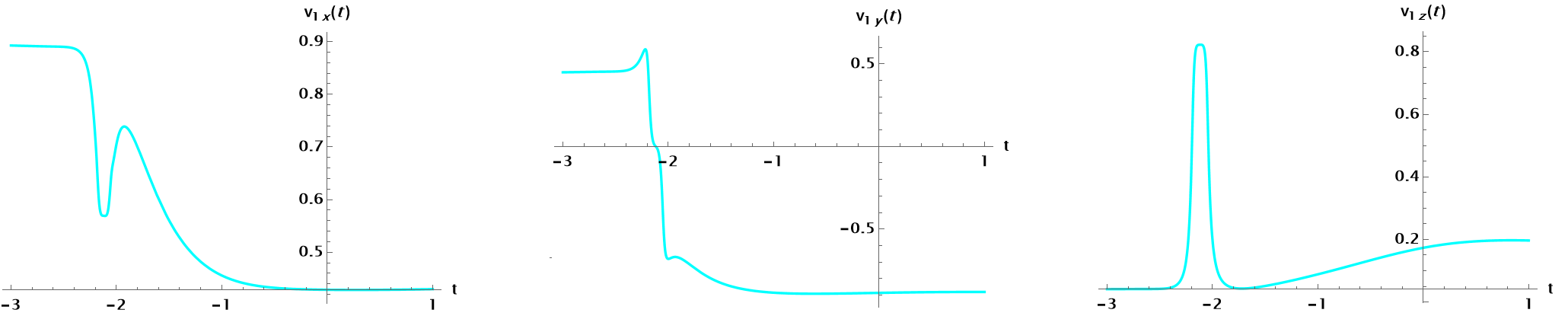}}
		\caption{Components $v_{1x}(t),v_{1y}(t),v_{1z}(t)$ of the eigenvector associated to the largest unstable eigenvalue $\lambda_1(t)$.}\label{fig:autovector1}
\end{figure}\\
Each direction $x,y,z$ in \autoref{fig:autovector1} should be identified, respectively, with $a,\dot a, \ddot a$. The plot reveals that the direction of maximum instability near the critical transition time $t_c$ satisfies $\dot a \approx 0$ and $\ddot a > 0 $, corresponding to an expansion around the reached critical time.

\section{Summary and discussion}
Quadratic gravity has attracted considerable attention, among others things, due to its good renormalization properties and its proper description of the inflationary era in the early Universe. In particular, it is a potential candidate to a Quantum Gravity theory which offers a nice UV behavior whereas reduces to GR at long distances, in spite o the fact that some issues such as causality or unitarity need to be further elaborated. 

In this paper we have studied the homogeneous and isotropic cosmological solutions of quadratic gravity. First we analyzed its field equations proving that the redundancy arising in ordinary Friedmann equations from GR also occurs in quadratic gravity, being necessary to fix the gauge in order to obtain a physical solution. Secondly, we have been able to reduce Stelle's action into a Starobinsky-like theory from a FLRW cosmological wievpoint, since the Weyl tensor vanishes for FLRW metrics. We have shown numerically that when $3\alpha>\beta$ (stable de Sitter branch) the Universe always presents a continuous expansion, while if $3\alpha<\beta$ (unstable de Sitter branch) the Universe is able to collapse unlike GR. Regarding the initial singularity, we have found solutions that have Big Bang followed by a continuous expansion ($3\alpha>\beta$) or a Big Crunch ($3\alpha<\beta$) , others that do not have Big Bang and end up collapsing to a Big Crunch ($3\alpha<\beta$) and finally some of them being singularity-free, not having either Big Bang nor Big Crunch ($3\alpha<\beta$). We performed a stability analysis of this last kind of solution by computing the second order variation of the action, showing that the transition from a non singular Universe to a singular one with emerging of a Big Bang is unstable. 

\section*{Acknowledgements}
We thank Diego Mu\~noz Zorrilla for fruitful discussions. M.A. is partially supported by Spanish MINECO/FEDER Grants No. PGC2022-126078NB-C21 funded by MCIN/AEI/ 10.13039/501100011033, ERDF A way of making Europe Grant; the Quantum Spain project of the QUANTUM ENIA of the {\sl Ministerio de Asuntos Econ\'omicos y Transformaci\'on Digital}, the {\sl Diputaci\'on General de Arag\'on Fondo Social Europeo (DGA-FSE)} Grant No. 2020-E21-17R of the Aragon Government, and {\sl Plan de Recuperaci\'on, Transformaci\'on y Resiliencia}- supported European Union – NextGenerationEU Program on {\sl Astrof{\'{\i}}sica y F{\'{\i}}sica de Altas Energ{\'{\i}}as}, CEFCA-CAPA-ITAINNOVA. M.P. is supported by {\sl Programa Investigo} funded by the European Union - NextGenerationEU {\sl Plan de Recuperaci\'on, Transformaci\'on y Resiliencia}.

\bibliographystyle{unsrturl}
\bibliography{bibliografia}
\nocite{*}



\end{document}